\documentclass{article}
\usepackage{graphicx}
\usepackage{amsmath}
\usepackage{multirow}
\usepackage{amssymb}

\title{Non-gaussian effects\\ in the cage dynamics of polymers}
\date{        }
\begin{document}

\maketitle

A. OTTOCHIAN $\dagger$$^{\ast}$\footnote{$^{\ast}$ Corresponding
   author. Email: ottochian@df.unipi.it}, C. DE
  MICHELE$\ddagger\S$ and D. LEPORINI$\dagger\ddagger$

\vspace{6pt}
  $\dagger${\em{Dipartimento di Fisica "Enrico Fermi", Universit\`{a} di Pisa, Largo B. Pontecorvo 3, I-56127 Pisa, Italy}};
  $\ddagger${\em{INFM-CRS Soft}}
  $\S$ { {\em Dipartimento di Fisica, Universit\`{a} di Roma "La
  Sapienza" Piazzale Aldo Moro, 2, 00185 Roma, Italy}}  

\begin{abstract}
The correlation between the fast cage dynamics and the structural relaxation is investigated in a model polymer system. It is shown that the cage vibration amplitude, as expressed by the Debye-Waller factor (DW), and the relaxation time $\tau_\alpha$ collapse on a single universal curve with a simple analytic form when the temperature, the density, the chain length and the monomer-monomer interaction potential are changed. For the physical states with the same $\tau_\alpha$ the coincidence of the mean-square displacement, the intermediate scattering function and the non-gaussian parameter is observed in a wide time window spanning from the ballistic regime to the onset of the Rouse dynamics driven by the chain connectivity. The role of the non-gaussian effects is discussed.  
\end{abstract}
\bigskip

\section{Introduction}
There is a growing interest in the relation between the fast vibrational dynamics and the long-time structural relaxation occurring in viscous systems and supercooled liquids close to their glass transition \cite{Angell95,Angell95B,HallWoly87,Dyre96,Dyre04,MarAngell01,Ngai00,Ngai04,StarrEtAl02,BordatNgai04,Harrowell06,BuchZorn92,ScopignoEtAl03,SokolPRL,Buchenau04,NoviSoko04,NovikovEtAl05,Johari06,Dyre06,OurNatPhys}.  This resulted in studies of the vibrational dynamics of both glasses \cite{MarAngell01,ScopignoEtAl03,SokolPRL,Buchenau04,NoviSoko04,NovikovEtAl05} and fluid systems \cite{BuchZorn92,StarrEtAl02,OurNatPhys}.

Recently, the universal correlation between the amplitude of the caged dynamics, as expressed by the  Debye-Waller factor (DW), and the structural relaxation time $\tau_\alpha$ has been evidenced by simulations and experiments on several physical systems including molecular liquids, polymers and metallic alloys in a wide range of fragility \cite{OurNatPhys}. It was found that the shape of the related scaling function  is also controlled by the non-gaussianity of the kinetic unit displacement \cite{OurNatPhys}. The present paper provides further insight on that issues by presenting new numerical results.

\section{Theory}
The glass transition has been pictured as 
 the freezing of a liquid in an Aperiodic Crystal Structure (ACS)
where the viscous flow is due to activated jumps over energy
barriers $\Delta E \propto k_B T a^2/ \langle u^2\rangle $ where $a$
is the displacement to overcome the barrier  and $\langle u^2\rangle$ 
is the DW factor of the liquid, i.e. the amplitude of the rattling motion within the cage of the nearest
neighbours atoms \cite{HallWoly87}.  The ACS picture leads to the Hall-Wolynes (HW) equation,
$\tau_\alpha, \eta \propto \exp (a^2/2\langle u^2\rangle )$, a relation which has been derived by others too \cite{BuchZorn92,Dyre04,NoviSoko03}. The HW equation relies on the condition that $\tau_\alpha$ exceeds the vibrational time scales. When the HW equation is compared with the experiments, one notes strong deviations from the predicted linear dependence between  $\log \tau_\alpha$ and $1/\langle u^2\rangle $ \cite{BuchZorn92}. To generalize the HW equation, one considers a distribution
$p(a^2)$ of the square displacement to overcome the energy barriers \cite{OurNatPhys}. 
Following the central limit theorem, the gaussian form is adopted,
$p(a^2)=\mathcal{N}\cdot
\exp [-{(a^2 - \overline{a^2})^2}/{2\sigma^2_{a^2}} ]$
where $\mathcal{N}$ is the normalization constant, $\overline{a^2}$
and $\sigma_{a^2}$ are the average and the standard deviation respectively. Note that, since  $\Delta E \propto a^2$, this choice corresponds to a gaussian distribution of energy barriers.
The distribution is taken to be independent of the state
parameters because the average displacement of the kinetic unit within $\tau_\alpha$ is weakly dependent on $\tau_\alpha$ \cite{Angell91}. In contrast, the DW factor depends on the state parameters \cite{Ngai04,Ngai00}. If one averages the HW expression over $p(a^2)$, one obtains  the following generalized HW expression (GHW):
\begin{equation}
\tau_\alpha = \tau_0 \exp\left( \frac{\overline{a^2}}{2\langle
u^2\rangle } + \frac{ \sigma^2_{a^2}}{8\langle u^2\rangle ^2 } \right ).
\label{parabola}
\end{equation}
Eq. \ref{parabola} neglects the very weak DW-dependence of $\tau_0$. If the linear temperature dependence of the DW factor is assumed, GHW  reduces to other results reported for supercooled liquids \cite{Bassler87}, polymers \cite{FerryEtAl53} or models of glassy relaxation \cite{MonthBouch96,EastArrow}.  

\section{Model}
A coarse-grained model of a linear freely-jointed polymer is used.  Non-bonded monomers
interact via a generalized Lennard-Jones pair potential $U_{p,q}(r)$ with $U_{p,q}(r)= \epsilon (q-p)^{-1} [ p (\sigma^\star/r)^q - q (\sigma^ \star/r)^p ]  + U_{cut}$ with $\sigma^\star = 2^{1/6} \sigma$. The parameters
$p$ and $q$ control the stiffness of the attractive and the repulsive part, respectively.
All quantities are in reduced units: length in units of $\sigma$, temperature in units of $\epsilon/ k_B$, and time in units of $\sigma \sqrt{m/\epsilon}$, where $m$ and $k_B$ are the monomer mass and the Boltzmann constant, respectively.  The energy unit is given by the depth of the potential well $\epsilon$. We also set $m = k_B = 1$.
The potential is cut and shifted to zero by $U_{cut}$ at $r=2.5$. 
The potential $U_{p,q}(r)$ reduces to the usual Lennard-Jones (LJ) potential by setting $p=6, q=12$ . Bonded monomers interact with a potential which is
the sum of the FENE (Finitely Extendible Nonlinear Elastic) potential and the LJ potential (see  ref. \cite{OurNatPhys} for further details). This results in a bond length $b=0.97$. Samples with $N \simeq 2000$ monomers were used.
Equilibration runs were performed in isothermal-isobaric (NPT) or canonical (NTV) ensembles. Data were collected under microcanonical conditions by integrating the equations of motion with a reversible multiple time steps algorithm, i.e. the r-RESPA algorithm \cite{tuckerman91}.
Physical states with different values of the temperature $T$, the density $\rho$, the chain length $M$ and the monomer-monomer interaction potential $U_{p,q}(r)$ were studied. See  ref. \cite{OurNatPhys} for further
details.

\section{Results and discussion}
In this section the translational dynamics and the relaxation of the
monomeric unit are studied. Changing the state parameters
$(T,\rho,M,p,q)$ results in changes of both the DW factor and the
relaxation time $\tau_\alpha$.  It was found that, when different
physical states have the same relaxation time,  both their
translational dynamics, as expressed by the mean square displacement
(MSD),  and their relaxation, as expressed by the self part of the
intermediate scattering function (ISF),  are coincident from the
ballistic regime up to the onset of the connectivity effects (Rouse
regime) at times fairly longer than $\tau_\alpha$
\cite{OurNatPhys}. The resulting clusters of curves are shown in
Fig.\ref{fig1} for both MSD ( $\langle r^{2}(t)\rangle
=N^{-1}\sum_{1}^{N}\langle ({\bf r}_i(t) - {\bf r}_i(0))^2 \rangle$ ) and ISF ( $F_s(q_{max},t) = N^{-1}
\sum_{1}^{N}\langle \exp[\i {\bf q}_{max}({\bf r}_i(t) - {\bf r}_i(0))] \rangle $, $q_{max}$ refers to the maximum of the static structure factor).
`[Insert figure 1 about here]'
Fig.\ref{fig1} also shows the definition of  $\tau_\alpha$ via the equation
$F_s(q_{max}, \tau_\alpha) =1/e$. The existence of clusters of physical states with similar dynamics over the wide time range from the vibrational regime to the long-time relaxation suggests that the latter are correlated. To investigate this issue, it was noted that  the location of the inflection point of the MSD does not depend on the state point, i.e. it always occurs at the same time  $t= t^\star$ in the log-log plot of
MSD \cite{OurNatPhys}. This gives the opportunity of a clear-cut definition of the DW factor as  $\langle
u^2 \rangle \equiv \langle r^2(t^\star) \rangle$. The plot of $\tau_\alpha$ vs $\langle
u^2 \rangle$ leads to a master curve for the numerical results well described by the GHW Eq.\ref{parabola} which fits the experimental results over about eighteen orders of magnitude \cite{OurNatPhys}.

We investigated if, in addition to MSD and ISF,  other quantities exhibit identical time-dependence when evaluated for the clusters of states with identical $\tau_\alpha$ values shown in Fig.\ref{fig1}. 
The results for  the non-gaussian parameter (NGF) $\alpha_2(t) =  (3\langle r^4(t)\rangle/5\langle r^2(t)\rangle^2) - 1$, which quantifies the dynamical heterogeneity of the system at a given time $t$ \cite{VogGlotz04},  are shown in Fig.\ref{fig2}. 
`[Insert figure 2 about here]'
NGF vanishes if the monomer displacement is a gaussian process. Fig.\ref{fig2} shows that NGF  increases between the end of the ballistic regime ($ t \sim 0.1$) and roughly $\tau_\alpha$.  It is apparent that, within the statistical uncertainty, physical states with coincident  MSD and ISF (see Fig.\ref{fig1}) have equal NGF up to about $\tau_\alpha$ too. For longer times the NGF  of physical states with equal $\tau_\alpha$ but different chain length differ also due to the onset of connectivity effects (Rouse regime) \cite{Smith01,Paul02}.

It has been already shown that the magnitude of the non-gaussian parameter is related to the curvature of the GHW Eq.\ref{parabola} when the plot $\log \tau_\alpha$ vs $\langle u^2 \rangle ^{-1}$ is considered \cite{OurNatPhys}. Indeed, the ratio of the quadratic and the linear terms of Eq.\ref{parabola} with respect to $\langle u^2 \rangle ^{-1}$,  $\mathcal{R} \equiv \sigma_{a^2}^2/(4\overline{a^2} \langle u^2 \rangle)$  increases with the height of NGF $\alpha_{2 \; max} $ and, if the latter vanishes, $\mathcal{R}$ does the same \cite{OurNatPhys}. The inset of Fig.\ref{fig2} shows the increase of $\alpha_{2 \; max}$ by increasing $\tau_\alpha$. If the non-gaussian effects are missing, ISF reduces to $F_s^g(q_{max},t)=\exp(-\frac{1}{6}q_{max}^2\langle r^2\rangle)$. The first correction to $F_s^g(q,t)$ due to the non-gaussian effects depends on NGF  and reads \cite{Nijboer1966} :
\begin{equation}
{\overline F_s}(q,t)= \exp \left (-\frac{1}{6}q^2\langle r^2 \rangle
\right ) \left [ 1 + \frac{1}{2}\alpha_2(t) \left(
\frac{1}{6}q^2\langle r^2\rangle \right) + \mathcal{O}\left ( \left(
\frac{1}{6}q^2\langle r^2\rangle \right )^2 \right ) \right ].
\label{Eq:ISFxNGF}
\end{equation}
`[Insert figure 3 about here]'
Fig.\ref{fig3} compares the numerical results for ISF with the
approximation ${\overline F}_s(q_{max},t)$. It is seen that, when
$\tau_\alpha \gtrsim 10^2$, ${\overline F}_s(q_{max},t)$ poorly
approximates the numerical results, thus showing that the non-gaussian
effects on the relaxation are not accounted for by the first
correction to $F_s^g(q_{max},t)$, i.e. they are not small. However, in
spite of the large deviations of ${\overline F}_s(q_{max},t)$ from the
exact results,  the inset of Fig.\ref{fig3} shows that the relative
error between $\tau_\alpha$ and the approximated estimate ${\overline
  \tau_\alpha}$ (to be defined by the equation ${\overline F}_s(q_{max},{ \overline \tau_\alpha} )= 1 / e$) is reasonable. 

\section{Conclusions}
The correlations between the fast dynamics of the monomers within the cage of the first neighbours and the long-time structural relaxation are studied. It is shown that physical states with equal $\tau_\alpha$ exhibit coincident MSD, ISF and NGF from the ballistic regime up to the onset of the connectivity effects (Rouse regime) at times fairly longer than $\tau_\alpha$. The first correction to the gaussian approximation of ISF disagrees from the numerical results for $\tau_\alpha \gtrsim 10^2$. However, the relative error between $\tau_\alpha$ and the approximated estimate ${\overline \tau_\alpha}$ stays within reasonable bounds. 
  
\section*{Acknowledgment}
Discussions with Dr. Luca Larini are acknowledged. Computational resources by ''Laboratorio per il Calcolo Scientifico''  (Physics Department, University of Pisa), financial support from MUR within the PRIN project ''Aging, fluctuation and response in out-of-equilibrium glassy systems'' and FIRB project "Nanopack" are acknowledged.

\newpage
\begin{figure}
  \begin{center}
     \caption{\label{fig1}(a) MSD time-dependence in selected cases. See ref.\cite{OurNatPhys} for details. The MSDs are multiplied by indicated factors.  (b) corresponding ISF curves. Four sets of clustered curves (A through D) show that, if states have equal $\tau_\alpha$ (marked with star on each curve ), the MSD and ISF curves coincide from times fairly longer than $\tau_\alpha$ down to the crossover to the ballistic regime at least. Adapted from ref.\cite{OurNatPhys}.}
  \end{center}
\end{figure}

\newpage
\begin{figure}
  \begin{center}
    \caption{ \label{fig2} Time-dependence of the the non-gaussian parameter (NGF) for the same states of Fig.\ref{fig1}. The stars denote the time $t =\tau_\alpha$. The plot shows that for states with equal  $\tau_\alpha$ not only MSD and ISF coincide between $t^{\star}$ and $\tau_\alpha$ (see Fig.\ref{fig1}) but also NGF does the same within the statistical uncertainty. The inset shows the increase of the maximum of NGF with $\tau_\alpha$. The dashed line is a parabolic guide for the eyes.  }
  \end{center}
\end{figure}
\vspace{10cm}
\newpage 
\begin{figure}
  \begin{center}
    \caption{ \label{fig3}Comparison between the numerical ISF (continuous lines) of Fig.\ref{fig1} and the first correction to the gaussian approximation ${\overline F}_s(q_{max},t)$ (dashed lines). The inset shows the relative error between $\tau_\alpha$ and the estimate $\overline \tau_\alpha$, as drawn from ${\overline F}_s(q_{max},{ \overline \tau_\alpha} )= 1 / e$. } 
  \end{center}
\end{figure}  
\clearpage
\begin{figure}
\vskip 3cm
\begin{center}
\includegraphics[width=1.0\linewidth]{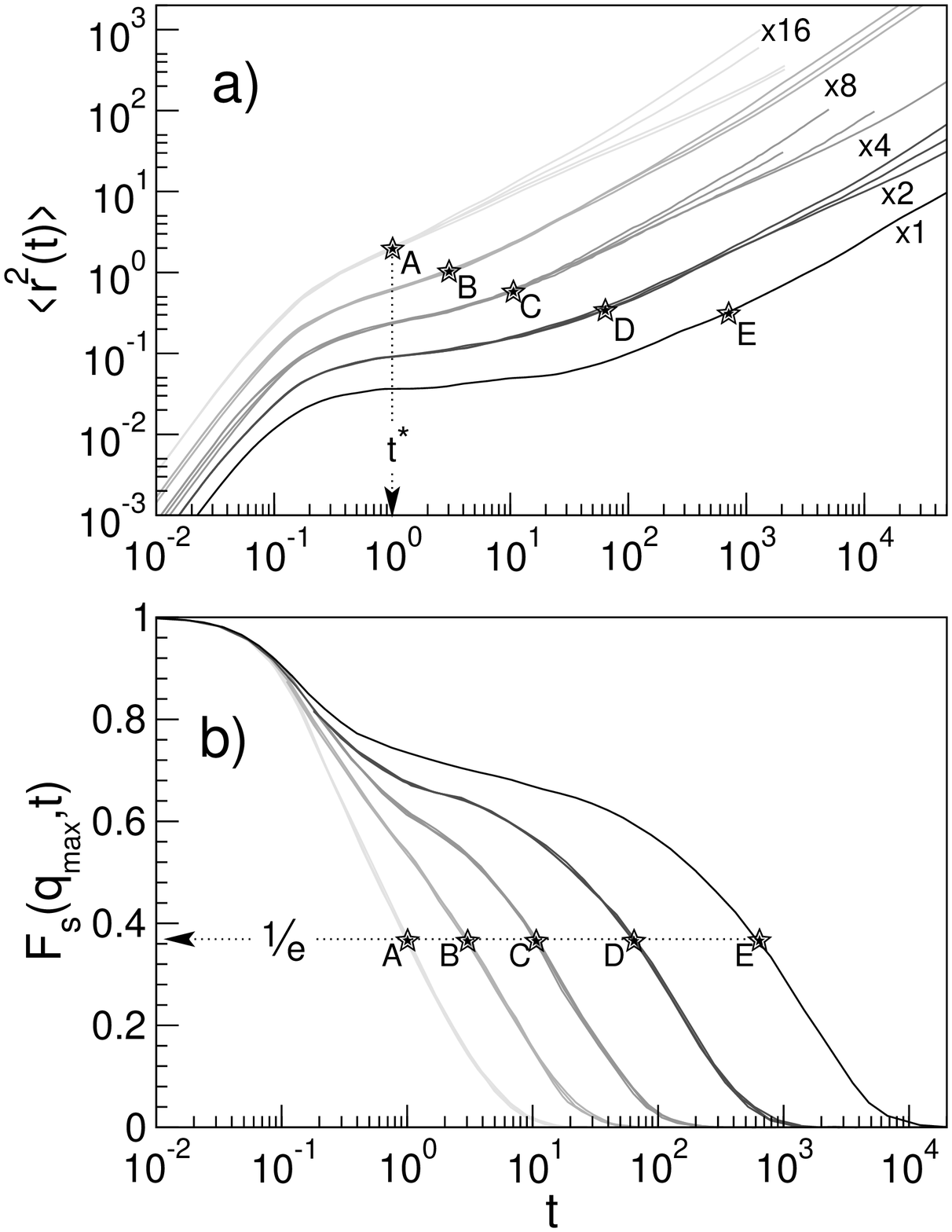}
\\
\vskip 1cm {\Large {\bf FIGURE 1}}
\end{center}
\end{figure}
\clearpage
\begin{figure}
\vskip 4.5cm
\begin{center}
\includegraphics[width=1.0\linewidth]{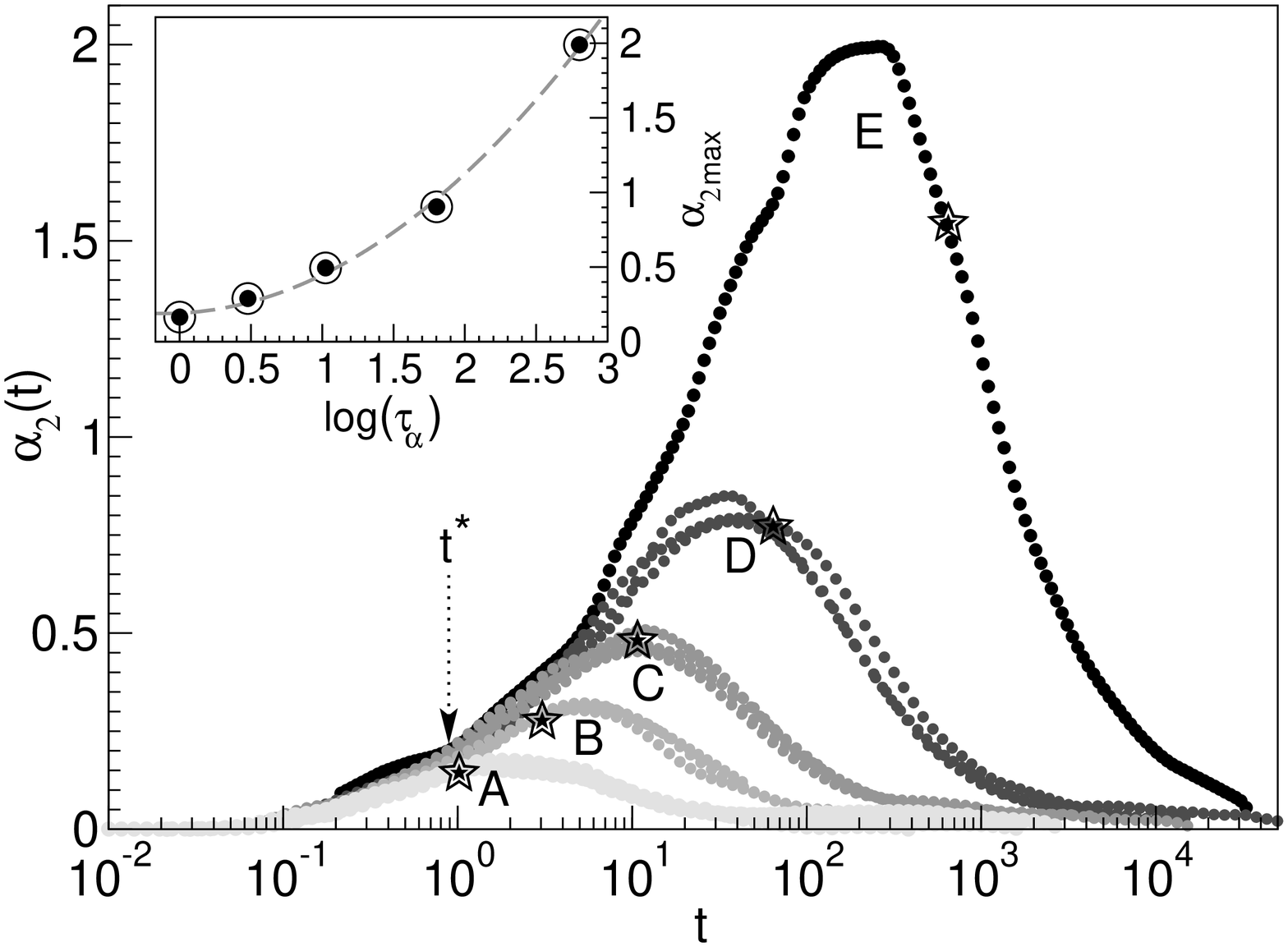}
\\
\vskip 1cm {\Large {\bf FIGURE 2}}
\end{center}
\end{figure}
\clearpage
\begin{figure}
\vskip 4.5cm
\begin{center}
\includegraphics[width=1.0\linewidth]{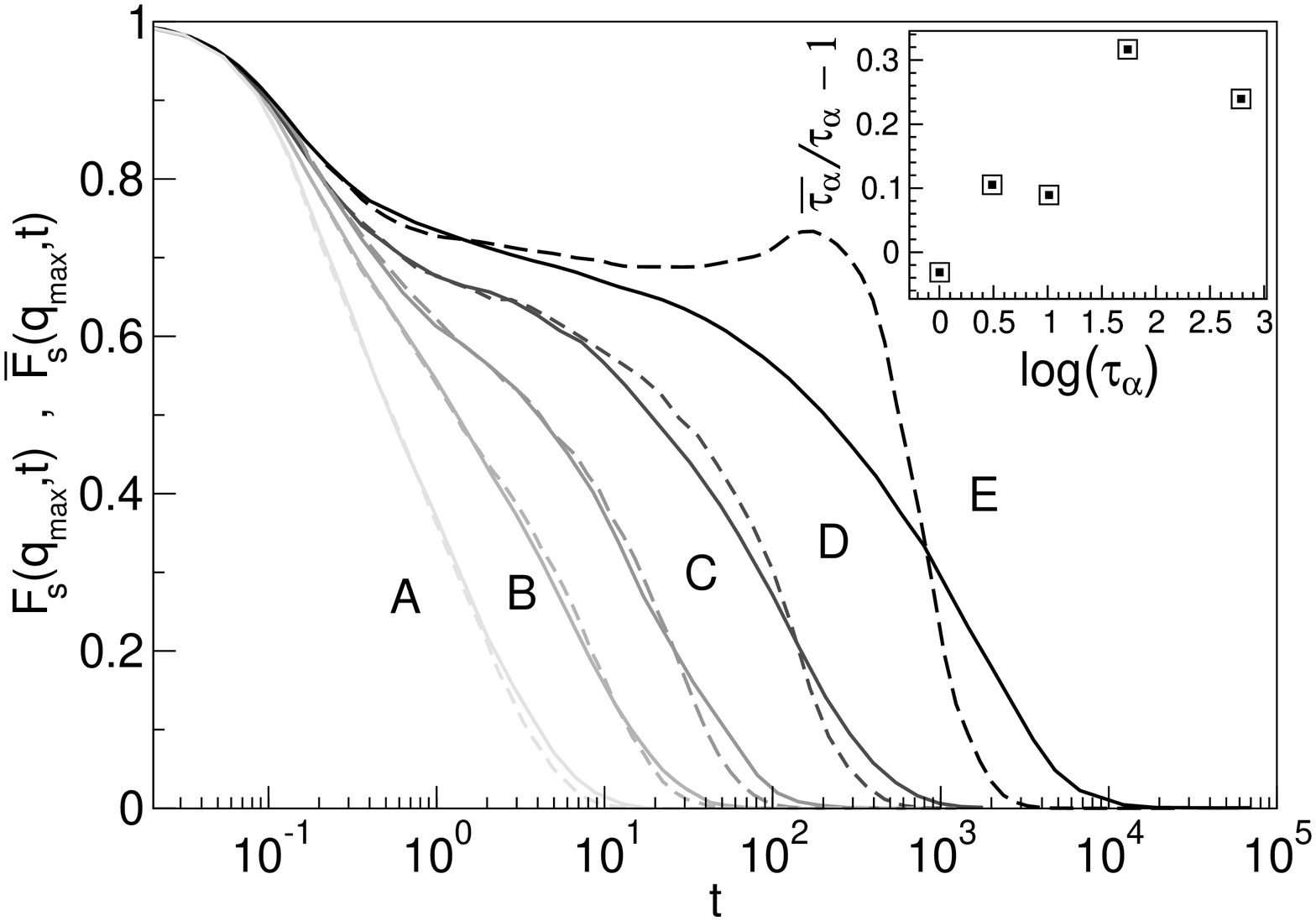}
\\
\vskip 1cm {\Large {\bf FIGURE 3}}
\end{center}
\end{figure}

\end{document}